\documentclass[preprint,12pt]{elsarticle}

\usepackage{xcolor}
\usepackage{soul}

\usepackage{amssymb}
\usepackage{upgreek}

\usepackage{tablefootnote}

\usepackage{sistyle}

\usepackage{setspace}
\journal{ }

\begin{document}

\newcommand{\mngesi}{Mn\textsubscript{5}(Si\textsubscript{x}Ge\textsubscript{1-x})\textsubscript{3} }
\newcommand{\mnge}{Mn\textsubscript{5}Ge\textsubscript{3} }
\newcommand{\mnsi}{Mn\textsubscript{5}Si\textsubscript{3} }

\begin{frontmatter}



\title{Epitaxial growth and magnetic properties of \mngesi thin films}


\author[CINAM]{Sueyeong Kang}
\author[CINAM]{Matthieu Petit\corref{cor1}}
\author[CINAM]{Vasile Heresanu}
\author[CINAM]{Alexandre Altié}
\author[polytech]{Thomas Beaujard}
\author[polytech]{Ganaël Bon}
\author[leeds]{Oscar Cespedes}
\author[leeds]{Brian Hickey}
\author[CINAM]{Lisa Michez}


\cortext[cor1]{Corresponding author}

\affiliation[CINAM]{organization={Aix-Marseille Univ, CNRS, CINaM UMR 7325},
            city={Marseille},
            country={France}}

\affiliation[polytech]{organization={Aix-Marseille Univ, Polytech Marseille, Dpt. of Materials Science and Engineering},
            city={Marseille},
            country={France}}

\affiliation[leeds]{organization={Condensed Matter group, University of Leeds},
            city={Leeds},
            country={United Kingdom}}




\begin{abstract}
Structural and magnetic properties of \mngesi thin films were investigated. Ferromagnetic \mnge and anti-ferromagnetic \mnsi thin films have been synthesized and characterized as these compounds exhibit interesting features for the development of spintronics. Here, \mngesi thin films were grown on Ge(111) substrates by co-deposition using molecular beam epitaxy. Crystalline thin films can be produced with controlled Si concentrations ranging from 0 to 1. The thin films were relaxed by dislocations at the interface with the substrate. A lattice parameter variation was observed as the Si content increased, which is comparable to previous works done in bulk. Reflection high-energy electron diffraction diagrams and X-ray diffraction profiles showed that lattice parameters a and c are shrinking and that the surface roughness and crystallinity degrade as the Si amount increases. Magnetometric measurements revealed a ferromagnetic behavior for all Si concentrations. The measured average ferromagnetic moment per manganese atom decreased from 2.33 to 0.05 $\upmu$\textsubscript{B}/Mn atom. No ferro to anti-ferromagnetic transition was observed contrary to the bulk \mngesi compound.
\end{abstract}


\begin{highlights}
\item Mn\textsubscript{5}(Si\textsubscript{x}Ge\textsubscript{1-x})\textsubscript{3} thin films are epitaxially grown on Ge(111) by MBE.
\item Epitaxial Mn\textsubscript{5}(Si\textsubscript{x}Ge\textsubscript{1-x})\textsubscript{3} thin films with a Si concentration up to x=0.6 are successfully synthesized.
\item A strong correlation is highlighted between the Si concentrations and the structural and magnetic properties of the thin films.
\end{highlights}


\end{frontmatter}


\section{Introduction}
\label{sec:Introduction}
 Manganese silicide and manganese germanide compounds have the advantage of being rare earth-free alloys. They are drawing great attention, both in the field of spintronics and magnetocaloric materials. Magnetic cooling, using the magnetocaloric effect (MCE), has a high potential as a solution for efficient thermal management. Manganite materials such as Mn-T-X (T = Ni, Co and X = Si, Ge) exhibit interesting values of the adiabatic temperature change ($\Delta T_{ad}$) and the magnetic entropy change ($\Delta S_{M}$) and room temperature MCE are obtained for some of these compounds \cite{kitanovski2020,chaudhary2019}.
 Among the manganite compounds, \mnge and \mnsi exhibit strong similarities but also fascinating differences, which are detailed in Table~\ref{table_1}. Both compounds can be integrated as thin films on commonly used substrates such as silicon or germanium.

\begin{table}[h!]
\centering
\begin{tabular}{ p{4cm} p{5.5cm} p{5.5cm}   } 
 \hline
  & \mnge & \mnsi  \\ 
 \hline
 Structure & \multicolumn{2}{c}{hexagonal D8$_{8}$ (P63/mcm)} \\
 
 Bulk lattice parameters (300 K) & a\textsuperscript{hex} = 7.18 $\angstrom$ \newline c = 5.05 $\angstrom$ \cite{castelliz1955} & a\textsuperscript{hex} = 6.91 $\angstrom$  \newline c = 4.81 $\angstrom$ \cite{aronsson1960}  \\

 Heat of formation (kJ.(mol of at)\textsuperscript{-1}) & -18.4 \cite{berche2014a} & -34.2 \cite{berche2014b} \\
 
 Epitaxial growth on: & Ge(111) \cite{zeng2003,petit2015} & Si(111) \cite{kounta2023} \\
 
 Epitaxial relationships & \mnge(0001)//Ge(111) \newline \mnge[$\overline{2}$110]//Ge[11$\overline{2}$] & \mnsi(0001)//Si(111) \newline \mnsi[$\overline{2}$110]//Si[11$\overline{2}$] \\
 
 Lattice mismatch with Ge(111) substrate & -3.75$\%$ & +0.27$\%$ \\
 
 Magnetic behaviors & metallic ferromagnet (FM) & metallic antiferromagnet (AFM) \\
 
 Relevant temperatures (K) & T$_{C}$ = 296 K \cite{tawara1963} & non-collinear AFM at T $\leq$ 66 K \newline collinear AFM for 66 K $\leq$ T $\leq$ 99 K \cite{brown1992,brown1995,biniskos2022}  \\
 
 Relevant features & uniaxial anisotropy along c-axis \cite{michez2015} & topological Hall effect for T $\leq$ 66 K \newline spontaneous Hall effect in epitaxial thin film for T $\leq$ 240 K \cite{kounta2023} \\ 

 Magnetocaloric properties & MCE effect, $\Delta S_{M}$ = 9.3 J.Kg\textsuperscript{-1}.K\textsuperscript{-1} (5T) \cite{tolinski2014} & inverse MCE effect, $\Delta S_{M}$ = 2 J.Kg\textsuperscript{-1}.K\textsuperscript{-1} (3T) \cite{gottschilch2012}\\
 
 \hline
\end{tabular}
\caption{Main properties of \mnge and \mnsi bulk compounds}
\label{table_1}
\end{table}

\begin{figure}[h!]
    \centering
    \includegraphics[width=90mm]{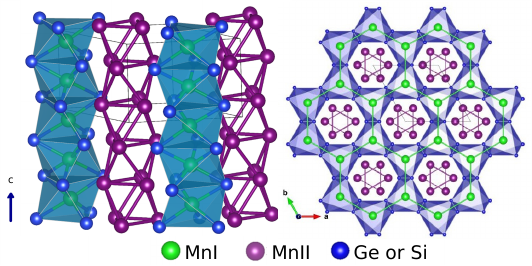}
    \caption{Mn\textsubscript{5}X\textsubscript{3} lattice structure with X = Ge or Si. Mn\textsubscript{I} atoms form layers of only Mn atoms perpendicular to the c axis. Some of the Mn\textsubscript{II} atoms create octahedra chains parallel to the c axis. (The diagrams presented here were generated using Vesta software \cite{vesta}.)  ~\label{lattice}}
\end{figure}

These two intermetallic alloys have a hexagonal D8\textsubscript{8}-type structure: the Mn atoms occupy two different types of sites. The first type of site is a layer of manganese atoms only (named Mn\textsubscript{I} in Wyckoff positions 4d (1/3, 2/3, 0)). The second type of site is a layer of manganese and germanium or silicon atoms (named Mn\textsubscript{II} on positions 6g (0.2358, 0, 1/4) plus column IV atoms on positions 6g (0.5992, 0, 1/4)). These two layers alternate as illustrated in Fig.~\ref{lattice}. The Mn\textsubscript{I} atoms form a rectilinear chain of metallic bonds, whereas the Mn\textsubscript{II} ones involve d\textsuperscript{2}s orbitals (sp\textsuperscript{2} for Ge or Si atoms) \cite{kanematsu1962}.  

The compounds can be grown by co-deposition epitaxy on Ge(111) substrate in the case of \mnge and on Si(111) substrate in the case of \mnsi ~\cite{petit2015,kounta2023}. In terms of magnetic properties, \mnge exhibits ferromagnetic behavior with a Curie temperature (T\textsubscript{C}) of 296 K, while \mnsi is considered to have antiferromagnetic characteristics with two first-order transitions associated with structural changes in bulk: below 66 K, \mnsi adopts a noncollinear antiferromagnetic phase (AF1). When the temperature exceeds this threshold, the first transition occurs, from AF1 to a collinear antiferromagnetic phase (AF2). The second transition takes place at T = 99 K, resulting in the transformation from the AF2 structure to a paramagnetic state~\cite{tawara1963,brown1992,brown1995,biniskos2022}.
The magnetocaloric properties of \mnge and \mnsi have also been studied. \mnge presents a second order magnetic phase transition and a value of $\Delta S_{M}$ equal to 9.3 J.Kg\textsuperscript{-1}.K\textsuperscript{-1} (5T) \cite{tolinski2014}. Polycristalline \mnsi exhibits an inverse magnetocaloric effect linked to the structural distorsion and $\Delta S_{M}$ = 2 J.Kg\textsuperscript{-1}.K\textsuperscript{-1} (3T, 62 K) \cite{gottschilch2012}.  
\\ 
Considering the great differences in the magnetic behaviors of \mnge and Mn\textsubscript{5}Si\textsubscript{3}, it is interesting to study the ternary alloys, denoted as Mn\textsubscript{5}(Si\textsubscript{x}Ge\textsubscript{1-x})\textsubscript{3}, with a variable silicon concentration (x) in the range [0; 1]. Previous research has predominantly dealt with either bulk \mngesi single crystals or polycrystalline samples, synthesized from melted pure Mn, Ge, and Si flakes, subsequently annealed several days at temperatures around 900-1120 K for homogenization~\cite{kappel1976,berche2015}. A comprehensive investigation into the structural, magnetic, and electrical properties of the bulk \mngesi alloys has been conducted. All bulk \mngesi compounds exhibit the same hexagonal D8$_{8}$ crystalline structure belonging to the space group P63/mcm at 300 K. The lattice parameters of the unit cell decrease with the increase of x. Berche \textit{et al} show the existence of a gradual kinetic phase separation phenomenon, leading to the transformation of the \mngesi solid solution into the separated \mnge and Mn\textsubscript{5}Si\textsubscript{3}, particularly with increasing annealing temperature~\cite{berche2015}. The \mngesi alloys exhibit a macroscopic ferromagnetic behavior within the x range of 0 to 0.75, with T\textsubscript{C} varying from 296 K to 151 K. The mean ferromagnetic moment per manganese atoms $\upmu_{F}$ decreases as x increases. At a Si concentration of approximately x = 0.8, the magnetic behavior tips from ferromagnetic to antiferromagnetic order, with total magnetization nearing zero at x = 0.85. Resistivity measurements conducted over a temperature range of [0; 30] K also confirm this transition~\cite{kappel1976,vancon1965,haul1979,zhao2006}. Theoretical calculations regarding Si substitution in \mnge indicate that the variation in the magnetic moments of the Mn\textsubscript{I} atoms is slightly greater than that of the Mn\textsubscript{II} atoms. This variation could be attributed to the modifications in the Mn-Mn distances with the Si content \cite{siberchicot1997}.
\\
In this article, we have employed a combination of techniques, including \textit{in situ} reflection high-energy electron diffraction (RHEED), X-ray diffraction (XRD), atomic force microscopy (AFM), and high-resolution transmission electronic microscopy (HR-TEM), to investigate the growth of \mngesi thin films on Ge(111) substrates via the molecular beam epitaxy (MBE) method. Additionally, the magnetic properties of the synthesized films were also characterized using a vibrating sample magnetometer (VSM) and a superconducting quantum interference device (SQUID). The purpose of our work is to provide a comparative analysis of \mngesi films in relation to the previous study on bulk with a view to their integration into a device heterostructure \cite{kappel1976,vancon1965,haul1979,zhao2006}.

\section{Experiments details}
\label{sec:Experiments details}
30 nm thick \mngesi thin films were prepared through the co-deposition of germanium, manganese, and silicon onto Ge(111) substrates using MBE. In the MBE setup with a base pressure better than 2 $\times$ 10\textsuperscript{-7} Pa, germanium and manganese were evaporated from conventional Knudsen effusion cells, while silicon was evaporated from a sublimation source, all sourced from MBE-Komponenten. Notably, this MBE cluster is equipped with \textit{in situ} RHEED, featuring a beam acceleration voltage of 30 kV, enabling real-time monitoring of the thin film growth process.

Before being introduced into the MBE chamber, Ge(111) substrates underwent chemical cleaning procedures \cite{mendez2008}. Within the ultra-high vacuum (UHV) environment, the substrates were pre-annealed at 720 K for several hours followed by a flash-annealing step reaching up to 1020 K, aimed at eliminating any residual germanium oxide present on the substrate surface. Subsequently, a 60 nm thick germanium buffer layer was grown over the Ge(111) substrate at 720 K, followed by annealing at 800 K to produce a high-quality germanium surface with a distinct c(2$\times$4) reconstruction, which was confirmed by \textit{in situ} RHEED observations.

Deposition rates of Mn, Ge, and Si were carefully calibrated using a quartz microbalance. The actual co-deposition onto the Ge buffer occurred simultaneously under UHV conditions maintained at 10\textsuperscript{-7} Pa and at a substrate temperature of 373 K. The presence of \mngesi layers was verified through the observation of the typical \mnge $(\sqrt{3} \times \sqrt{3})$R30$^{\circ}$ RHEED pattern \cite{zeng2004,mendez2008}. No subsequent annealing was performed on the thin films after the co-deposition in order to avoid inter-diffusion between the film and the substrate.

To assess crystalline orientation and quality of the thin films, 2-dimensional X-ray diffraction (2D-XRD) diagrams were acquired using a high brilliancy rotating anode Rigaku RU-200BH, equipped with an image plate detector Mar345 and operating with the non-monochromatic Cu K$\upalpha$ radiation ($\uplambda = 1.54180 ~\angstrom$). The experimental resolution is about 0.3$^{\circ}$ and the diffraction angle is varied within the range of 5 to 35 degrees to cover a 2$\uptheta$ range spanning from 10 to 70 degrees. The diffraction profiles plotting the X-ray diffracted intensities over 2$\uptheta$ degrees are generated by integrating the intensities of 2D diffraction diagrams for an equal radial distance.
The quality of interfaces and their correlation with epitaxial growth on the Ge(111) surface was investigated through HR-TEM measurements. These measurements were performed using a JEM-2100F (JEOL) instrument, operating at an accelerating voltage of 200 kV and a spatial resolution of 2.3 $\angstrom$. Prior to HR-TEM analysis, samples underwent thinning via a precision ion polishing system (PIPS), allowing the acquisition of cross-sectional images.
Surface topographies were obtained by AFM with a Nanoscope IIIA Multimode from Digital
instruments on a 2 x 2 $\upmu$m\textsuperscript{2} area.
The magnetic properties of the thin films were probed using a Maglab 9T VSM from Oxford instruments and an MPMSXL SQUID magnetometer from Quantum Design.

\section{Results and discussion}
\label{sec:Results and discussion}
\subsection{Structure of the \mngesi thin films}
\begin{figure}[h!]
    \centering
    \includegraphics[width=90mm]{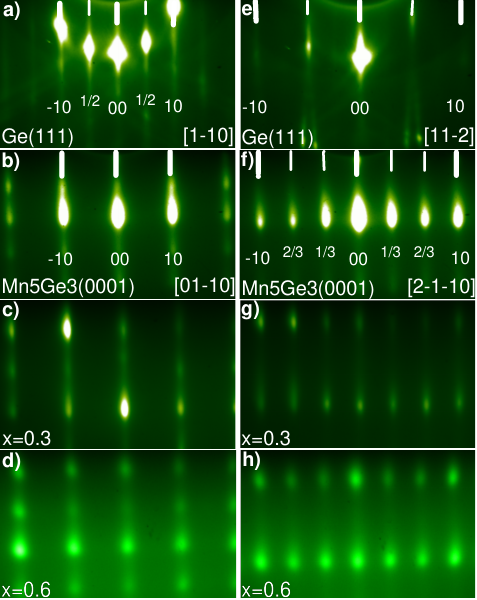}
    \caption{RHEED patterns taken along the Ge(111)-[1$\overline{1}$0] a)-d) and Ge(111)-[11$\overline{2}$] e)-h) azimuths. The bulk 0$\times$0 and 1$\times$1 streaks are indicated with large white rods and the reconstructed streaks are marked with small white rods. Patterns a) and e): Ge(111)c(2$\times$4) surface prior to the Ge, Si, and Mn co-deposition. Patterns b)-d) and f)-h): RHEED patterns taken at the end of the co-deposition of 30 nm thick \mngesi films with a silicon concentration x equal to 0 (i.e. \mnge film, patterns b) and f)), x = 0.3 (Mn\textsubscript{5}(Si\textsubscript{0.3}Ge\textsubscript{0.7})\textsubscript{3} film, patterns c) and g)), and x = 0.6 (Mn\textsubscript{5}(Si\textsubscript{0.6}Ge\textsubscript{0.4})\textsubscript{3} film, patterns d) and h)). (The electron beam intensity is not constant throughout the RHEED screenshots)
    ~\label{rheed pattern}}
\end{figure}

A series of \mngesi samples with a variation of x from 0 to 1 were grown on Ge(111) substrates by co-deposition with a careful adjustment of the values of the Ge, Si, and Mn flux. Epitaxial growth was performed by simultaneously opening the shutters of Ge, Si, and Mn cells. The RHEED patterns of the initial Ge(111) surface exhibited a well defined c(2$\times$4) reconstruction (Fig.~\ref{rheed pattern} a) and e)), indicating a clean and long scale ordered surface.

\begin{figure}[h!]
    \centering
    \includegraphics[width=90mm]{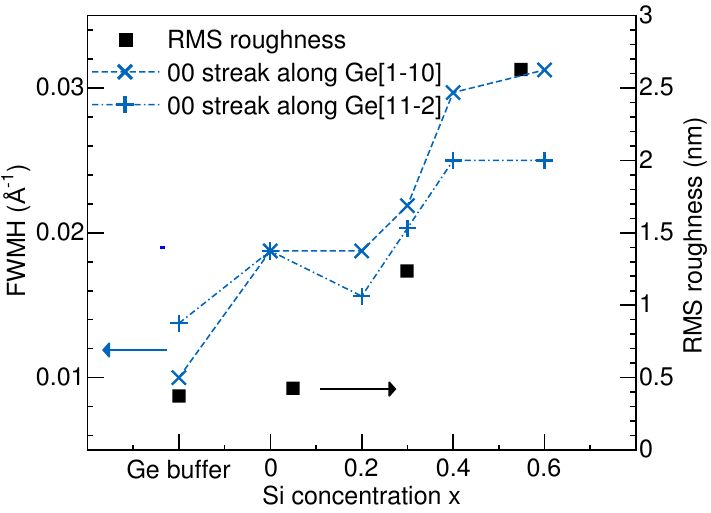}
    \caption{left Y-axis: FWMH of the 00 streaks measured on RHEED patterns of the surface of 30 nm thick \mngesi thin films, along the Ge[1$\overline{1}$0]-Mn\textsubscript{5}(Si\textsubscript{x}Ge\textsubscript{1-x})\textsubscript{3}[01$\overline{1}$0] and Ge[11$\overline{2}$]-Mn\textsubscript{5}(Si\textsubscript{x}Ge\textsubscript{1-x})\textsubscript{3}[2$\overline{1}$$\overline{1}$0] azimuths. Right Y-axis: RMS roughness of three \mngesi thin films (x = 0.05, 0.3, and 0.55) measured by AFM.
    ~\label{roughness}}
\end{figure}

The identical \mnge characteristic $(\sqrt{3} \times \sqrt{3})$R30$^{\circ}$ reconstruction patterns were observed through RHEED analysis conducted on \mngesi thin films, indicating the fact that the surface structure remained unchanged regardless of the Si concentration (Fig.\ref{rheed pattern} b)-d) and f)-h)). However, as the Si substitution increased, RHEED patterns became spottier and the streaks got more blurred. To quantify this observation, Fig.~\ref{roughness} displays the evolution of the full width at half maximum (FWHM) of the 00 RHEED streaks over x. The FWHM values of each 00 streak were obtained by plotting an intensity profile perpendicular to the streaks on RHEED patterns recorded at the end of the co-deposition process for 30 nm thick \mngesi films. This parameter is mostly linked to the vertical amplitude of roughness of the given surface~\cite{lagally1988,chevrier1991}. Additionally, the root mean square (RMS) roughnesses were evaluated by AFM on some of these films in agreement with RHEED observations. 

\begin{figure}[ht]
    \centering
 \begin{tabular}{r}

\includegraphics[width=90mm]{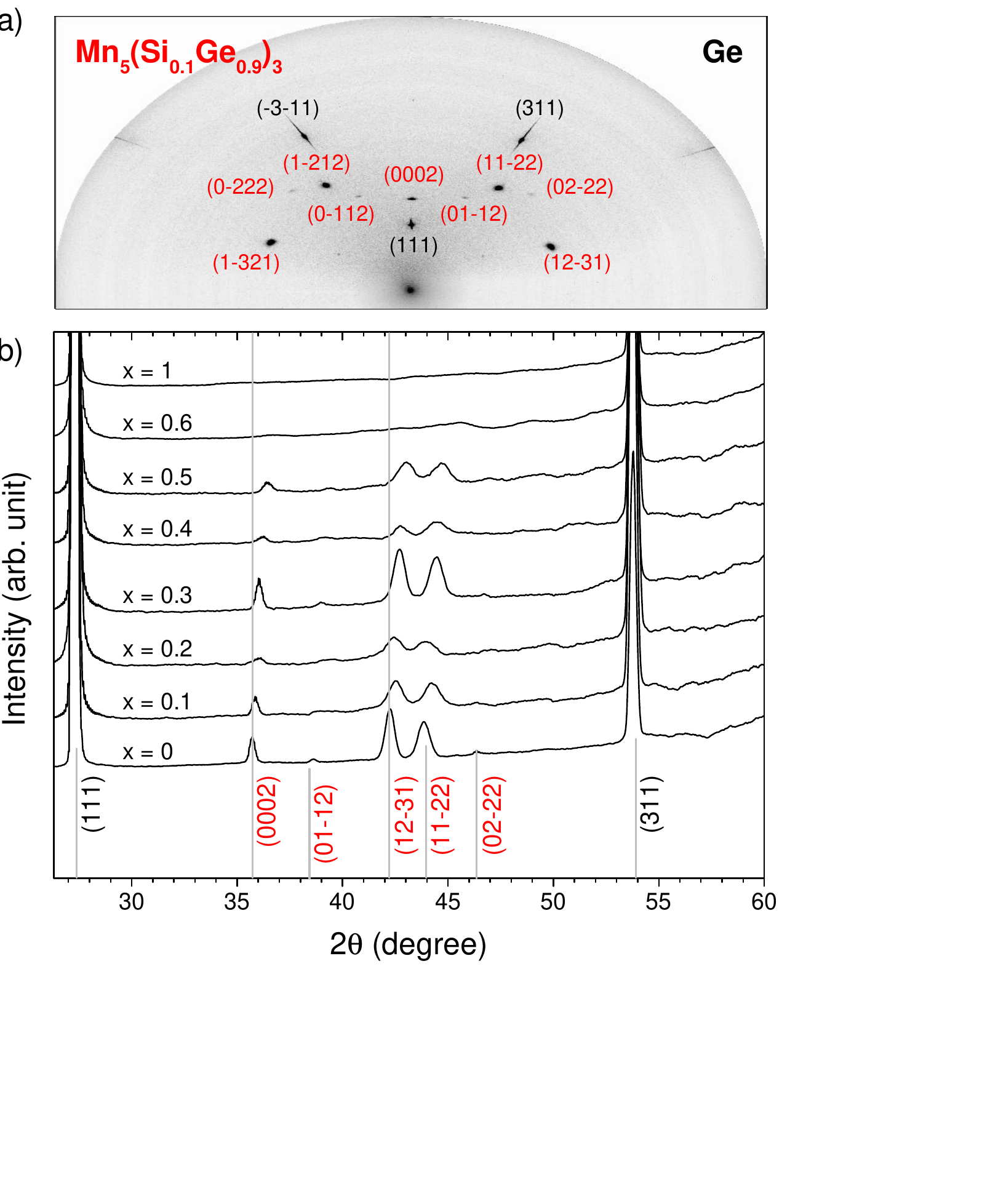} \\
\includegraphics[width=90mm]{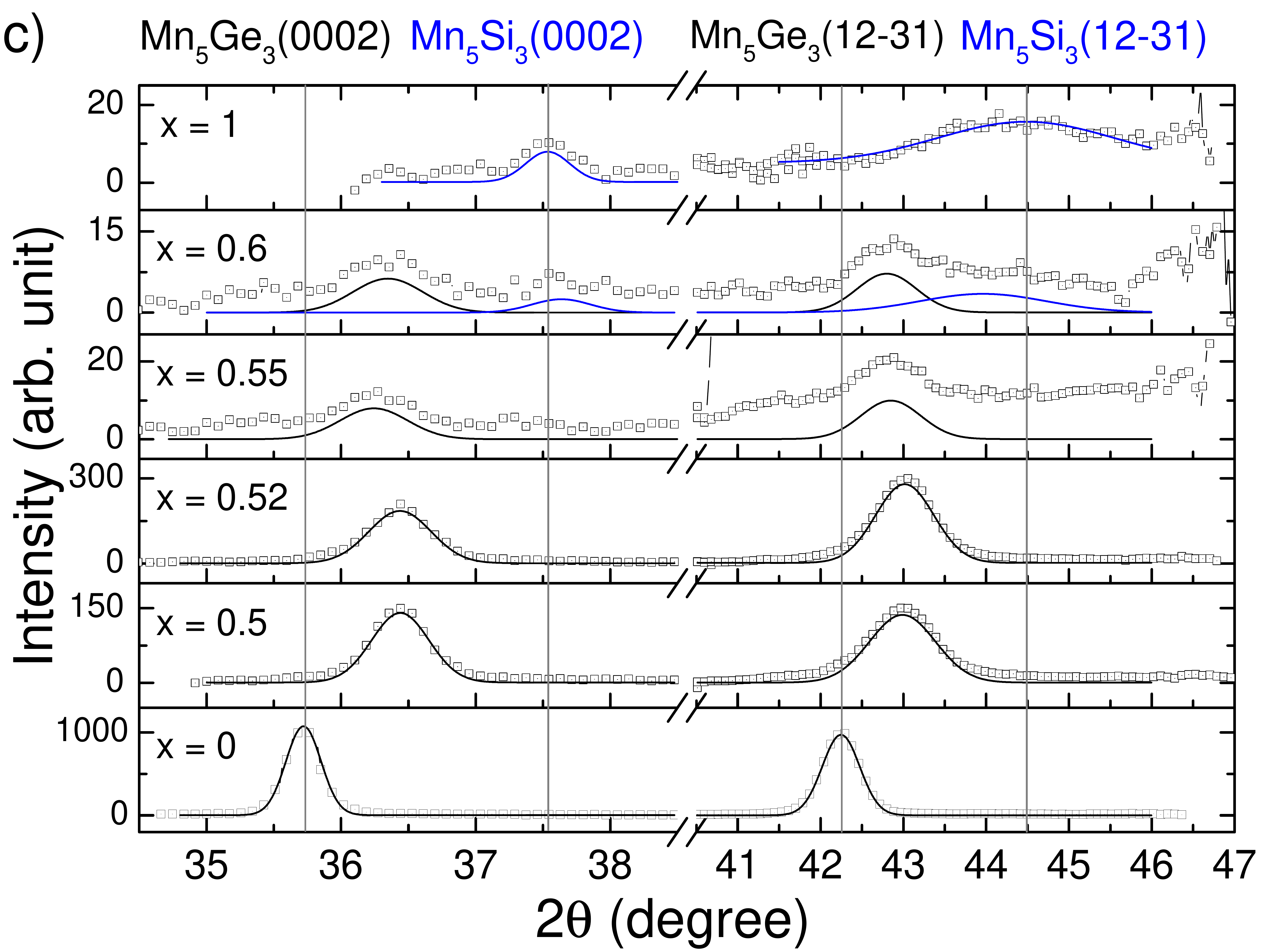}

 \end{tabular}
    \caption{X-ray diffraction data of \mngesi samples. a) A typical 2D-XRD map obtained for Mn\textsubscript{5}(Si\textsubscript{0.1}Ge\textsubscript{0.9})\textsubscript{3}, with each spot corresponding to the diffraction angles of the labeled planes.  b) XRD profiles generated from the integrated intensities for equal radial distance, i.e. for equal 2$\uptheta$ angle, of the 2D-XRD maps of \mngesi samples with Si concentration x = 0 to 1. As these profiles are an integration of 2D maps, they contain both symmetric and asymmetric diffraction peaks. The lattice planes associated with the deflection angles of Germanium and \mngesi are marked in black and red, respectively. c) XRD profiles focusing on high Silicon content thin films, ranging from x = 0.5 to x = 1. Symbols indicate the experimental data points and solid lines represent Gaussian fits of the data. Powder diffraction angles for \mnge and \mnsi are indicated by gray lines. ~\label{xrd1}}
\end{figure}

To assess the crystallinity of the \mngesi films, 2D-XRD maps were recorded for various Si concentrations. Fig.~\ref{xrd1} a) shows a representative diffraction map of a Mn\textsubscript{5}(Si\textsubscript{0.1}Ge\textsubscript{0.9})\textsubscript{3} film.

\begin{figure}[h!]
    \centering
    \includegraphics[width=140mm]{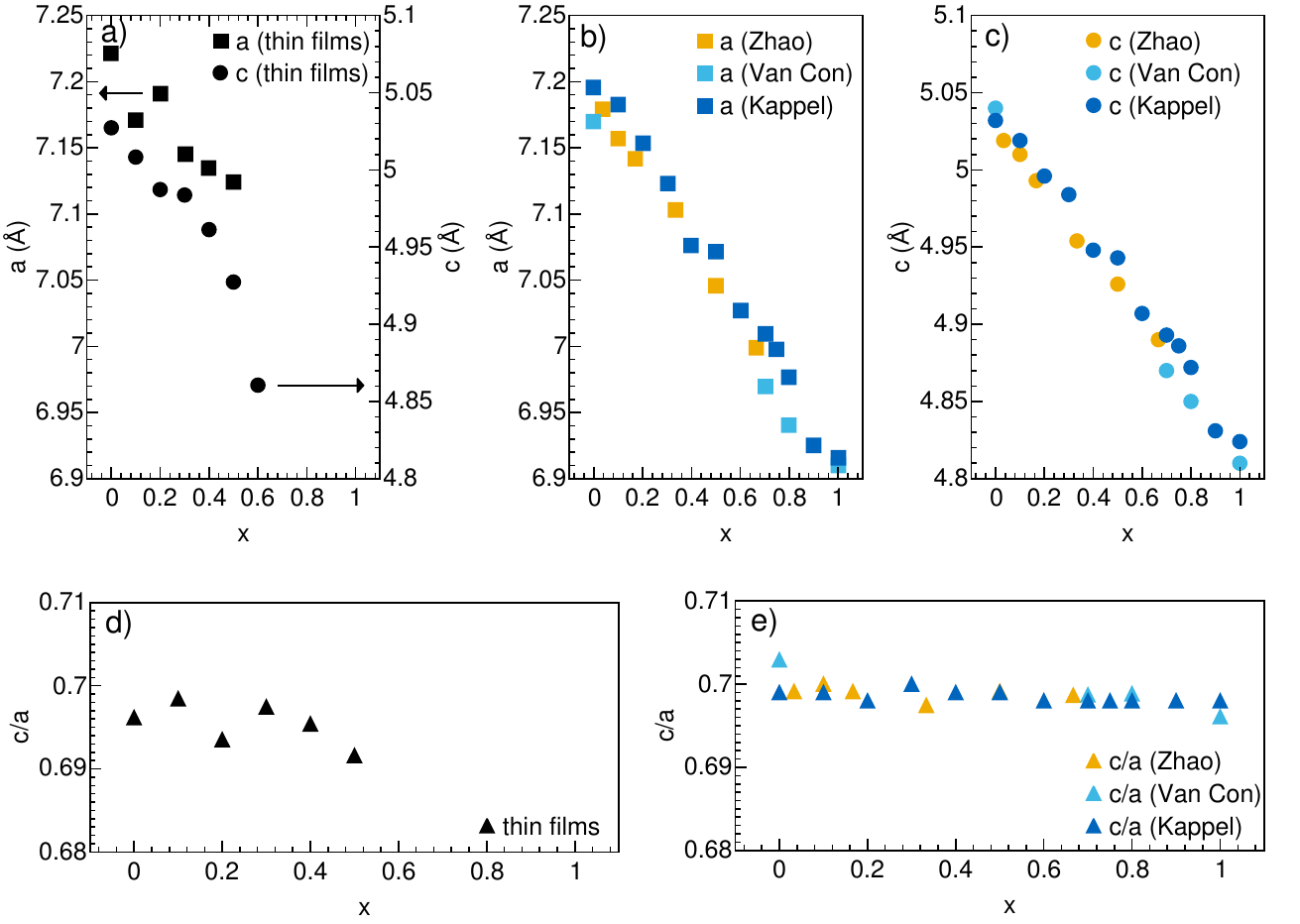}
    \caption{a) variation of the lattice parameters a and c of the \mngesi thin films versus the Si concentration x. b) and c) variation of the lattice parameters a and c respectively versus x in bulk samples, from ref.\cite{kappel1976,vancon1965,zhao2006}. d) and e) variation of the ratio c/a in the thin films and bulk compounds, respectively.  
    ~\label{lattice_para}}
\end{figure}

Fig.~\ref{xrd1} b) presents the plots of the integrated intensities of the 2D-XRD maps of thin films with Si concentrations ranging from x = 0 to 1. The two intense peaks at 2$\uptheta$ = 27.30$^{\circ}$ and 53.73$^{\circ}$ are originated from the Ge substrate and correspond to Ge(111) and Ge(311) planes, respectively. Peaks around 2$\uptheta$ = 35.48$^{\circ}$, 42.44$^{\circ}$ and 43.76$^{\circ}$ can be attributed to the (0002), (12$\overline{3}$1), and (11$\overline{2}$2) diffraction planes of Mn\textsubscript{5}(Si\textsubscript{x}Ge\textsubscript{1-x})\textsubscript{3}, with a slight shift towards higher 2$\uptheta$ angles at increasing Si concentrations. Above x = 0.5 (Fig.~\ref{xrd1} c)), a significant drop in the diffracted intensities as well as broadening of the peaks are observed. In addition, a new diffraction peak appears around 2$\uptheta$ = 37.5$^{\circ}$, which corresponds to the powder diffraction of Mn\textsubscript{5}Si\textsubscript{3}(0002) peak. This indicates the possible existence of \mnsi crystalline grains in the films. As Si atoms are expected to substitute Ge atoms, the \mngesi films are predicted to crystallize in the hexagonal D8\textsubscript{8}-type structure, with the lattice parameters falling between those of \mnge and \mnsi compounds. Since phase separation has been demonstrated in bulk Mn\textsubscript{5}(Si\textsubscript{x}Ge\textsubscript{1-x})\textsubscript{3}, the formation of \mnsi grains in the thin films can be favorable during the co-deposition at higher Si concentrations rather than forming a \mngesi film with a sole and homogeneous Si concentration. The decrease in the peak intensities and their broadening combined with the evolution of the RHEED patterns (Fig.~\ref{rheed pattern}) show that the crystalline quality of the \mngesi films dropped with the increase of x. Regarding the peak positions, an increase in Si substitution x leads to a peak shift towards a higher 2$\uptheta$ angle, indicating a shrinkage of the lattice parameters. The lattice parameters a and c were calculated based on two diffraction peaks, (0002) and (12$\overline{3}$1). Fig.~\ref{lattice_para} a) and d) present the evolution of both a and c, as well as the ratio c/a, in relation to Si concentrations below  0.6. Indeed, the accuracy of determining parameters is affected by the low intensity of the diffraction peaks above x = 0.6. However, both a and c values decrease linearly as x increases within the accessible concentration range and align well with those of the bulk alloys found in literature, as shown in Fig.~\ref{lattice_para} b), c), and e)~\cite{vancon1965,kappel1976,zhao2006}.   

\begin{figure}[h!]
    \centering
    \includegraphics[width=90mm]{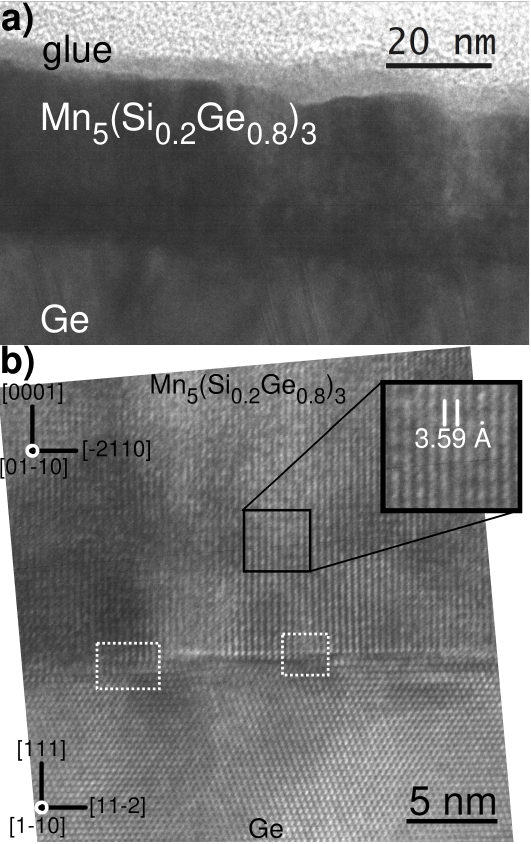}
    \caption{a) Transmission electron microscopy (TEM) image $\sim$30 nm thick Mn\textsubscript{5}(Si\textsubscript{0.2}Ge\textsubscript{0.8})\textsubscript{3} film. b) High-resolution TEM image focused on the Mn\textsubscript{5}(Si\textsubscript{0.2}Ge\textsubscript{0.8})\textsubscript{3}/Ge(111) interface with a zoom (delimited with a black line) on the Mn\textsubscript{5}(Si\textsubscript{0.2}Ge\textsubscript{0.8})\textsubscript{3} phase. The centers of interfacial dislocations are highlighted in white dotted squares.~\label{tem}}
\end{figure}

Fig.~\ref{tem} a) displays a cross-sectional HR-TEM image of a 30 nm thick Mn\textsubscript{5}(Si\textsubscript{0.2}Ge\textsubscript{0.8})\textsubscript{3} thin film on a Ge(111) substrate. The surface of the film exhibits some roughness. Fig.~\ref{tem} b) is focused on the interface between the thin film and the substrate and unveils a crystalline film epitaxially grown on the Ge(111) substrate. The identified zone axis of the film is [01$\overline1$0] and the [0001] axis is parallel to the Ge[111] in accordance with the epitaxial relationships established using RHEED and XRD techniques. The lattice parameter calculated from the TEM image is 7.18 $\angstrom$, which is consistent with the value found by XRD (7.19 $\angstrom$).   

\subsection{Mismatch accommodation}
\begin{figure}[h!]
    \centering
    \includegraphics[width=90mm]{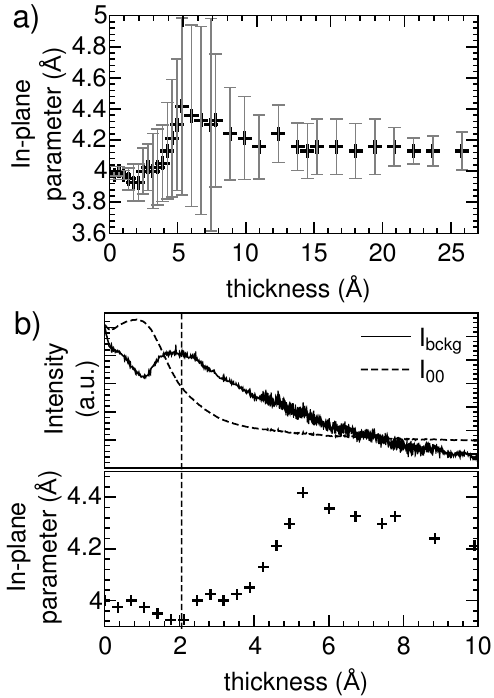}
    \caption{a) Tracking of the in-plane lattice parameter over the first 26 $\angstrom$ of the co-deposition obtained by measuring the RHEED streaks spacing in the Ge(111)-[1$\overline{1}$0] azimuth in the case of a Mn\textsubscript{5}(Si\textsubscript{0.3}Ge\textsubscript{0.7})\textsubscript{3} thin film. b) Zoom over the first 10 $\angstrom$ of the co-deposition of (top) I\textsubscript{(00)} the intensity of the 00 streak in the Ge(111)-[1$\overline{1}$0] azimuth and I\textsubscript{bckg} the intensity of the background of the RHEED movie and (bottom) the in-plane parameter. ~\label{rheed_intensity}}
\end{figure}

The fact that the lattice parameters a and c of the \mngesi films are close to those in the bulk (Fig.~\ref{lattice_para}) means that the thin films are relaxed. To get a better insight into the relaxation process, we recorded a movie of the RHEED diagram during the first 26 $\angstrom$ of the co-deposition growth in the Ge(111)-[1$\overline{1}$0] azimuth. From this movie, we extracted the evolution of the in-plane lattice parameter of the growing film and the evolution of the intensity of the 00 streak (I\textsubscript{00}) and of the background intensity (I\textsubscript{bckg}) over the thickness of the film. The in-plane lattice parameter was calculated by converting the spacing of the RHEED streaks into a real space distance. The background intensity was measured at a location between the 00 and 01 streaks. Fig.~\ref{rheed_intensity} a) displays the evolution of the in-plane parameter over the whole first 26 $\angstrom$ of the co-deposition of a Mn\textsubscript{5}(Si\textsubscript{0.3}Ge\textsubscript{0.7})\textsubscript{3} thin film as an illustration. It is representative of the other concentrations. 

Starting at 4.00 $\angstrom$, this in-plane parameter value at the given azimuth corresponds to the lattice parameter of the Ge(111) surface, a\textsubscript{Ge}/$\sqrt{2}$. Between 1 and 12 $\angstrom$ of thickness, the value underwent an abrupt change. The longer error bars reflect an increase in the blurring of the RHEED stripes. Then, it stabilized at 4.13 $\angstrom$ around a film thickness of 12 $\angstrom$. Along the observed \mngesi(0001)-[01$\overline{1}$0] azimuth, the spacing of the RHEED streaks corresponds to a\textsubscript{Mn\textsubscript{5}(Si\textsubscript{x}Ge\textsubscript{1-x})\textsubscript{3}}/$\sqrt{3}$ which gives a\textsubscript{Mn\textsubscript{5}(Si\textsubscript{0.3}Ge\textsubscript{0.7})\textsubscript{3}} = 7.15 $\angstrom$, in good agreement with the XRD results (Fig.~\ref{lattice_para} a)). From this evolution of the in-plane parameter, we concluded that the thin film had relaxed in this short span of thickness. Looking more precisely in the range of 0 to 10 $\angstrom$ of co-deposition (Fig.~\ref{rheed_intensity} b)), the change in the in-plane parameter consisted of a slight decrease followed by an increase up to 4.40 $\angstrom$ again followed by a final decrease to the steady state value of 4.13 $\angstrom$. Furthermore, the RHEED intensities exhibited abrupt fluctuations, particularly when I\textsubscript{bckg} increased before decreasing and becoming stable. This increase is indicative of a transient disordering of the surface at the given co-deposition thickness. Closer examination of the TEM image (Fig.~\ref{tem} b)) showed that the interface exhibited some stacking faults (marked by the dotted white square on the right side of Fig.~\ref{tem} b)) and misfit dislocations (marked by the dotted white square on the left side of Fig.~\ref{tem} b)). The formation of these defects explains the variation of the in-plane parameter measured by RHEED and the increase of the background intensity and has already been observed in the case of the growth of \mnge thin films \cite{padova2011,petit2019}. The accommodation of the lattice mismatch between the \mngesi films and the Ge(111) substrate seems to take place in a very thin layer of less than two \mngesi lattices thick. Note that after this transitory phase, I\textsubscript{00} also reached a steady state, indicating that the growth front has stabilized.

\subsection{Discussion of the \mngesi growth}
The crystalline quality and the surface roughness of the \mngesi thin films degrade as x increases. This is not intuitive given the heats of formation and the lattice mismatches of the compounds (Table \ref{table_1}). Based on these parameters, \mnsi appears to be the most favorable compound for epitaxial growth on Ge(111). Several other phenomena can be invoked to understand the evolution of \mngesi films with x.
First, the low interface energy of the \mnge/ Ge(111) system ($\upgamma_{Mn_{5}Ge_{3}/Ge(111)}$ = 0.53 J.m$^{-2}$) plays a role in the epitaxial stability of the germanide phase on the Ge(111) substrate~\cite{arras2010,guerboukha2022}. The growth of \mnsi films on Si(111) requires a MnSi interfacial layer to reduce the surface energy and thus promote the crystallization of \mnsi \cite{kounta2023}. Although the surface energies of Ge(111) and Si(111) are not the same ($\upgamma_{Ge(111)}$ = 1.30 J.m$^{-2}$,  $\upgamma_{Si(111)}$ = 1.74 J.m$^{-2}$), the interfacial energy of the \mngesi/ Ge(111) system may not offer favorable conditions for the film nucleation as x increases~\cite{stekolnikov2002}.
Next, the initial stages of Mn atom adsorption on the Ge(111) surface are believed to be important for the crystal growth of \mnge films. An impinging Mn atom first takes up position on a Ge adatom site (in a T\textsubscript{4} site) before shifting to a neighboring H\textsubscript{3} adsorption site \cite{zeng2004}. Two out of three H\textsubscript{3} sites are occupied by Mn atoms. These well-defined Mn positions initiate the Mn\textsubscript{I} rectilinear chains of the \mnge lattice with the c-axis perpendicular to the substrate surface \cite{padova2011}. In the case of Mn\textsubscript{5}(Si\textsubscript{x}Ge\textsubscript{1-x})\textsubscript{3}, Si atoms are expected to substitute for Ge atoms, but lower surface mobility of Si atoms than Ge atoms on Ge(111) surface and potential competition for adsorption sites between silicon and manganese atoms may be detrimental to the further ordering of the \mngesi layer \cite{maree1987,zhachuk2016}.
Finally, although co-deposition (and not solid phase epitaxy) is considered in this article, diffusion phenomena may play a significant role in the formation and crystallization of the \mngesi films. Mn diffusion has been proven to be vacancy mediated and quite fast during phase formation in \mnge \cite{portavoce2012,abbes2012}, and the chains of vacancies created parallel to the Mn\textsubscript{I} rectilinear chains above the unoccupied H\textsubscript{3} sites can provide preferential diffusion pathways. Moreover, \mnge thin films can be synthesized on Ge(111) substrates with good crystallinity without an annealing step after co-deposition, whereas crystalline \mnsi films are more difficult to produce on Si(111) substrates, as they require annealing at 500 K \cite{petit2015,kounta2023}. Noting also that the \mngesi lattice shrinks with increasing x, which may slow diffusion through the free volumes available in the lattice, the formation of \mngesi crystalline films is hampered by increasing Si concentration.

\subsection{SQUID-VSM magnetometry}
\begin{figure}[h!]
    \centering
    \includegraphics[width=90mm]{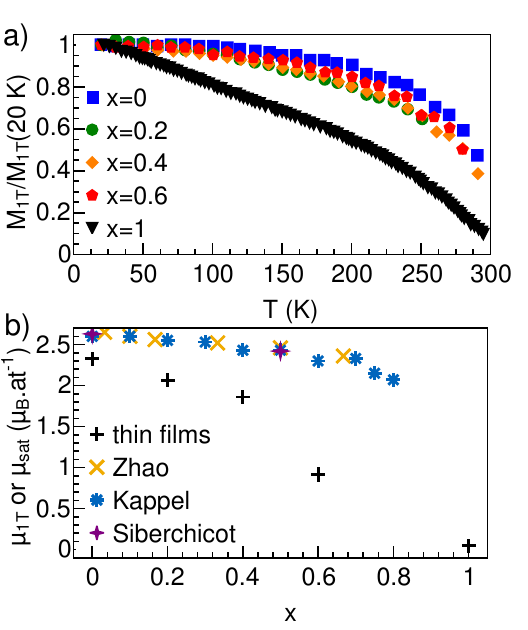}
    \caption{a) Temperature dependence of the magnetization (M\textsubscript{1T}) of \mngesi samples obtained by VSM(x = 0 to 0.6) and SQUID(x = 1) recorded in a 1 T-field. b) Evolution of the average saturation magnetic moment $\upmu$\textsubscript{sat} at 20 K versus Si concentration x. (Kappel, Siberchicot and Zhao are data from ref.\cite{kappel1976,siberchicot1997,zhao2006})} ~\label{m1T}
\end{figure}

The magnetic behavior of \mngesi samples was measured using VSM (x = 0, 0.2, 0.4, and 0.6) and SQUID (x = 1). Magnetic hysteresis loops were acquired by VSM up to a magnetic field of 1 T, covering temperatures ranging from 20 K to 290 K with 10 K intervals. The raw data were processed to minimize experimental contributions: the magnetic moments of germanium substrates were subtracted. For x = 1 sample, magnetization versus temperature data were collected at a magnetic field of 1 T within the temperature range of 2 K to 400 K using SQUID, due to the signal intensity being too weak for a VSM measurement. For comparison of the magnetic moment recorded at 1 T (M\textsubscript{1T}) with respect to silicon content, the measured M\textsubscript{1T} were normalized to the volume of the samples and are plotted on Fig.~\ref{m1T}~a). At 1 T, the saturation value is reached for the magnetization for the samples with low values of x (M\textsubscript{1T} = M\textsubscript{sat}). 

Overall, magnetizations decrease with increasing temperature, corresponding to the demagnetization process of ferromagnetic materials. \mnsi on Ge(111) shows an unusual M-T curve with \textit{a piori} ferromagnetic characteristics. The T\textsubscript{C} of samples with x = 0 to 0.6 were determined by fitting the M-T curves using a phenomenological model that obeys the Bloch power law at low temperature and reproduces the critical behavior near T\textsubscript{C} \cite{kuzmin2005}:

\begin{equation}
M(T) = M(0) \left[ 1 - s \left( \frac{T}{T_{C}} \right) ^{\frac{3}{2}} - \left( 1 - s \right) \left( \frac{T}{T_{C}} \right)^{p} \right] ^ {\frac{1}{3}}
\end{equation}

where $s \geq 0$ and $p \geq \frac{3}{2}$ are fitting parameters. For the x = 1  and even for x = 0.6 samples, the shapes of the M-T curves are not well described by the model as they do not exhibit a sharp transition around T\textsubscript{C}. Yasasun et al. also observed an indistinct and broad transition near T\textsubscript{C} in the M-T curve of their \mnsi dominated sample~\cite{gunduz2019effect}. The magnetic transition temperature of the \mnsi sample was therefore evaluated using the derivative $\frac{dM}{dT}$ of the SQUID experimental data. All the T\textsubscript{C} are reported on Table \ref{table_2} and decrease as x increases.

\begin{table}[h!]
\centering
    \begin{tabular}{ p{2.5cm} p{1.5cm} p{1cm}  } 
    \hline
    Si content (x) & T\textsubscript{C} (K) & R\textsuperscript{2}  \\ 
    \hline
    x = 0.0 & 310 & 0.94 \\
    x = 0.2 & 311 & 0.97 \\
    x = 0.4 & 300 & 0.99 \\
    x = 0.6 & 296 & 0.70 \\
    x = 1.0 & 285$\pm$4 & - \\
    \hline
    \end{tabular}    
\caption{Curie temperatures (T\textsubscript{C}) obtained by fitting the experimental data of Fig.~\ref{m1T} with the phenomenological model from ref.\cite{kuzmin2005}. R\textsuperscript{2} is the correlation coefficient. T\textsubscript{C} of x = 1 is obtained using SQUID ZFC-FC measurement. }
\label{table_2}
\end{table}

The temperature of the magnetic transition of \mnsi on Ge(111) substrate is different from the temperature found on \mnsi films grown on Si(111) substrate~\cite{kounta2023}.
Additionally, the mean magnetic moments per Mn atoms $\upmu$\textsubscript{1T} were calculated at 20 K for each Si content. Fig.~\ref{m1T} b) displays their evolution as the silicon content increases. $\upmu$\textsubscript{1T} decreases with a drop at x $\geq$ 0.4. In particular, at x = 1 $\upmu$\textsubscript{1T} is equal to 0.05 $\upmu$\textsubscript{B} (M\textsubscript{1T} = 19.4 emu$\cdot$cm\textsuperscript{-3}) at 20 K, which significantly contrasts with the value at x = 0.6 where it reaches $\upmu$\textsubscript{1T} = 0.92 $\upmu$\textsubscript{B} (M\textsubscript{1T} = 376.2 emu$\cdot$cm\textsuperscript{-3}) at 20 K. Comparing with the data from the bulk, the values of the mean magnetic moments per Mn atoms at saturation $\upmu$\textsubscript{sat} also exhibit a decrease but the slope is not as high as in thin films \cite{kappel1976,siberchicot1997,zhao2006}. A ferromagnetic to antiferromagnetic transition is not observed within the limits of silicon concentrations in thin films accessible with the current growth method, whereas in bulk samples, a transition is observed around x = 0.8 \cite{kappel1976,vancon1965,haul1979,zhao2006}. The slightly apparent ferromagnetic behavior of the \mnsi film suggests the presence of very low crystalline \mngesi compounds at the interface with the Ge(111) substrate.

\section{Conclusion}
\label{sec:Conclusion}
\mngesi thin films with x ranging from 0 to 1 were fabricated using MBE by co-deposition of Mn, Ge, and Si. Both structural and magnetic properties were investigated using RHEED, XRD, AFM, TEM, VSM, and SQUID. RHEED and XRD technics reveal that the lattice structure of thin films remains hexagonal D8$_{8}$ (P63/mcm) structure alike to \mnge regardless of Si concentrations. By comparing the series of XRD integrated profiles and their peak shifts, we can conclude that both a and c parameters of hexagonal lattice shrink by increasing Si concentration. This result is in good agreement with the previous study on the bulk medium. The degradation of the crystalline and surface quality of the films is observed on higher Si concentration of x $\geq$ 0.5. Notably, \mnsi(0002) and \mnsi(12$\overline{3}$1) diffraction peaks were observed at this condition, indicating the occurrence of a possible phase separation. The analysis of the live-time RHEED patterns and TEM images shows that \mngesi thin films on Ge(111) are almost fully relaxed and that the relaxation takes place through dislocations in the interface. VSM and SQUID investigation display the decline of the overall magnetization as the temperature increases, confirming that \mngesi thin films present similar magnetic behavior as \mnge thin film. Surprisingly, the \mnsi film still exhibits a very weak ferromagnetic behavior. The mean magnetic moments per Mn atoms is affected by the increase of the Si content. However, the origin of this decrease is unclear and the substitution of Ge atoms by Si atoms as well as the increase in the crystalline disorder in the films may be an important factor. Further optimization of the growth process is required to synthesize crystalline \mngesi thin films over the entire Si concentration, and additional nuclear magnetic resonance (NMR) studies may provide a deeper understanding of the films' magnetic behavior.

\section*{Acknowledgments}
\label{sec:Acknowledgments}
The authors would like to thank Cyril Coudreau and Christopher Genelot for technical support on the MBE cluster, and Alain Ranguis for the AFM images. 
This article is based upon work from COST-OPERA Action CA20116, supported by COST (European Cooperation in Science and Technology).


\bibliographystyle{elsarticle-num} 
\bibliography{mngesi}





\end{document}